\documentclass[journal=jacsat,manuscript=article]{achemso}

\usepackage{chemformula} 
\usepackage[T1]{fontenc} 

\author{Titiksha Srivastava}
\affiliation{Univ. Grenoble Alpes, CEA, CNRS, Grenoble INP*, INAC-Spintec, 38000 Grenoble, France, * Institute of Engineering Univ. Grenoble Alpes}%

\author{Marine Schott}
\affiliation{Univ. Grenoble Alpes, CEA, CNRS, Grenoble INP*, INAC-Spintec, 38000 Grenoble, France, * Institute of Engineering Univ. Grenoble Alpes}%
\affiliation{Univ. Grenoble Alpes, CNRS,  N\'eel Institute, F-38042 Grenoble, France}

\author{Rom\'eo Juge}
\affiliation{Univ. Grenoble Alpes, CEA, CNRS, Grenoble INP*, INAC-Spintec, 38000 Grenoble, France, * Institute of Engineering Univ. Grenoble Alpes}%

\author{Viola K\v{r}i\v{z}\'akov\'a}
\affiliation{Univ. Grenoble Alpes, CNRS,  N\'eel Institute, F-38042 Grenoble, France}

\author{ Mohamed Belmeguenai}
\affiliation{Laboratoire des Sciences des Proc\'ed\'es et des Mat\'eriaux, Univ. Paris 13 Nord,  93430 Villetaneuse, France}

\author{Yves Roussign\'e}
\affiliation{Laboratoire des Sciences des Proc\'ed\'es et des Mat\'eriaux, Univ. Paris 13 Nord,  93430 Villetaneuse, France}%

\author{Anne Bernand-Mantel}
\affiliation{Univ. Grenoble Alpes, CNRS,  N\'eel Institute, F-38042 Grenoble, France}

\author{Laurent Ranno}
\affiliation{Univ. Grenoble Alpes, CNRS,  N\'eel Institute, F-38042 Grenoble, France}

\author{Stefania Pizzini}
\affiliation{Univ. Grenoble Alpes, CNRS,  N\'eel Institute, F-38042 Grenoble, France}

\author{Salim-Mourad Ch\'erif}
\affiliation{Laboratoire des Sciences des Proc\'ed\'es et des Mat\'eriaux, Univ. Paris 13 Nord,  93430 Villetaneuse, France}

\author{Andrey Stashkevich}
\affiliation{Laboratoire des Sciences des Proc\'ed\'es et des Mat\'eriaux, Univ. Paris 13 Nord,  93430 Villetaneuse, France}

\author{St\'ephane Auffret}
\affiliation{Univ. Grenoble Alpes, CEA, CNRS, Grenoble INP*, INAC-Spintec, 38000 Grenoble, France, * Institute of Engineering Univ. Grenoble Alpes}

\author{Olivier Boulle}
\affiliation{Univ. Grenoble Alpes, CEA, CNRS, Grenoble INP*, INAC-Spintec, 38000 Grenoble, France, * Institute of Engineering Univ. Grenoble Alpes}

\author{Gilles Gaudin}
\affiliation{Univ. Grenoble Alpes, CEA, CNRS, Grenoble INP*, INAC-Spintec, 38000 Grenoble, France, * Institute of Engineering Univ. Grenoble Alpes}

\author{Mairbek Chshiev}
\affiliation{Univ. Grenoble Alpes, CEA, CNRS, Grenoble INP*, INAC-Spintec, 38000 Grenoble, France, * Institute of Engineering Univ. Grenoble Alpes}

\author{Claire Baraduc}
\affiliation{Univ. Grenoble Alpes, CEA, CNRS, Grenoble INP*, INAC-Spintec, 38000 Grenoble, France, * Institute of Engineering Univ. Grenoble Alpes}

\author{H\'el\`ene B\'ea}
\affiliation{Univ. Grenoble Alpes, CEA, CNRS, Grenoble INP*, INAC-Spintec, 38000 Grenoble, France, * Institute of Engineering Univ. Grenoble Alpes}
\email{helene.bea@cea.fr}

\title{Large Voltage Tuning of Dzyaloshinskii-Moriya Interaction: a Route towards Dynamic Control of Skyrmion Chirality}


\begin{document}

\begin{tocentry}

	\includegraphics[width=\linewidth]{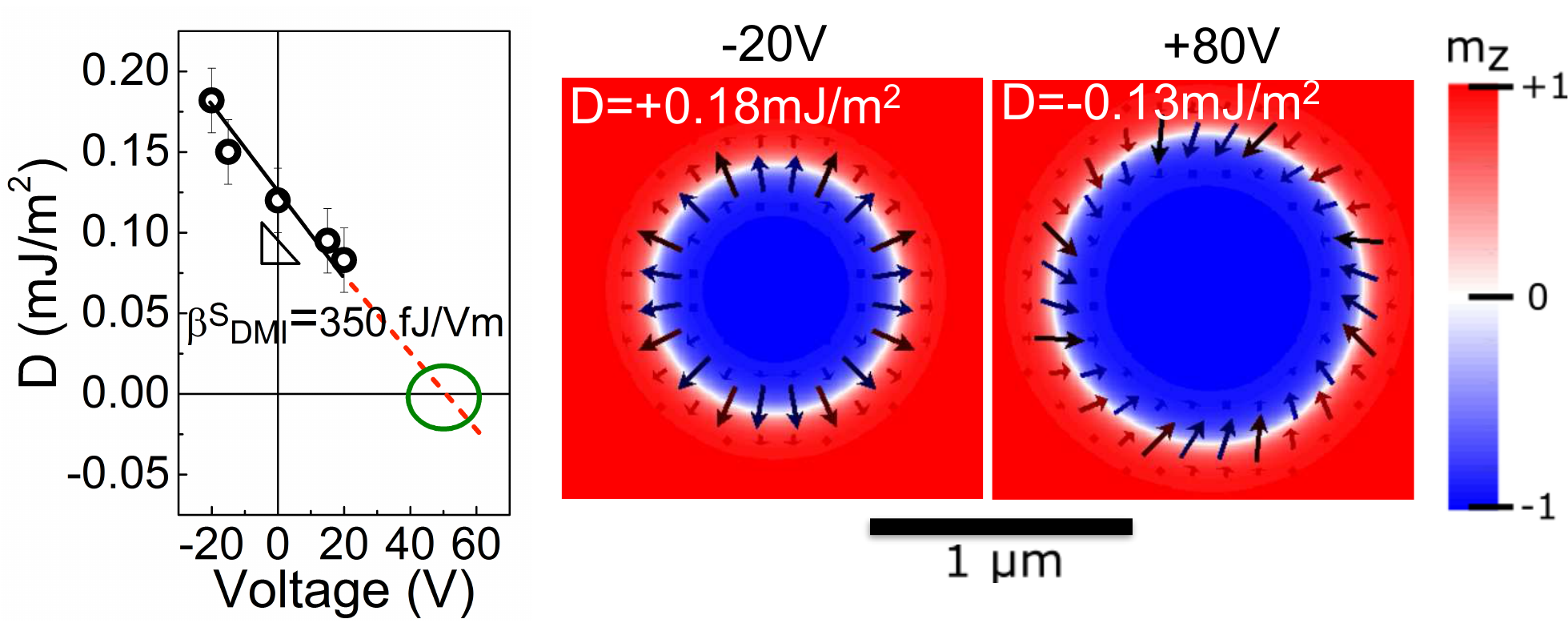}

\end{tocentry}

\begin{abstract}
Electric control of magnetism is a prerequisite for efficient and low power spintronic devices. More specifically, in heavy metal/ ferromagnet/ insulator heterostructures, voltage gating has been shown to locally and dynamically tune magnetic properties like interface anisotropy and saturation magnetization. However, its effect on interfacial Dzyaloshinskii-Moriya Interaction (DMI), which is crucial for the stability of magnetic skyrmions, has been challenging to achieve and has not been reported yet for ultrathin films. Here, we demonstrate 130\% variation of DMI with electric field in Ta/FeCoB/TaO$_x$ trilayers through Brillouin Light Spectroscopy (BLS). Using polar-Magneto-Optical-Kerr-Effect microscopy, we further show a monotonic variation of DMI and skyrmionic bubble size with electric field, with an unprecedented efficiency. We anticipate through our observations that a sign reversal of DMI with electric field is possible, leading to a chirality switch. This dynamic manipulation of DMI establishes an additional degree of control to engineer programmable skyrmion based memory or logic devices.

\end{abstract}


Topologically non trivial magnetic structures called skyrmions\cite{BogdV} were initially discovered in bulk crystals like MnSi and FeGe at low temperature \cite{Muhl,YuX}.  They have recently been observed at room temperature in  thin  trilayer systems consisting of a heavy metal (HM), a ferromagnet (FM) and an insulator (I) like Ta/FeCoB/TaOx  \cite{Jiang}, Pt/CoFeB/MgO \cite{Woo} and Pt/Co/MgO \cite{Boulle}. Broken inversion symmetry and spin orbit coupling in these trilayers lead to antisymmetric exchange called interfacial Dzyaloshinskii-Moriya Interaction \cite{Robler,Nagaosa}(DMI) which gives rise to non-collinear magnetic alignments. The sign of the DMI determines the chirality of N\'eel walls. It thus plays a fundamental role in stabilizing skyrmions of the preferred chirality, along with dipolar, exchange and interface anisotropy energies. 
In the HM/FM/I systems current induced skyrmion motion and their local manipulation with current through spin orbit torque has been demonstrated, which is very promising for spintronic applications \cite{Jiang,Woo}. Moreover, skyrmions can be locally and dynamically controlled by electric field through voltage gating \cite{HsuEfEsw,Marine}. This effect could advantageously be used in low power devices, as demonstrated in other spintronics devices \cite{TsymbalEFE,MatsuEFE}. Owing to the short screening length in metals, electric field mainly affects the FM/I interface. It induces a rapid charge re-distribution and thus an immediate modification of interfacial magnetic properties. However for long duration application of voltage, additional phenomena such as ion migration or charge trapping may occur, which produce slower but usually stronger effects \cite{DienyMair, Bauer}.
  
The recent experimental demonstrations of  electric field induced creation and annihilation of skyrmions at room temperature \cite{HsuEfEsw,Marine} has been  mainly attributed to a tuning of the interface anisotropy and saturation magnetization\cite{Marine, Upad, HsuEfEsw, KangVCS, NakEFEsky}. There have been some speculations on a possible DMI variation with voltage, but its observation is cumbersome. Although a very small voltage induced variation of DMI in thick FM films (20nm) consisting of epitaxial Au/Fe/MgO by K. Nawaoka et al.\cite{NawaEFDMI} was reported, DMI variation  with voltage in ultrathin films has not been addressed yet. 
In fact, in   previous studies,  DMI was attributed to the HM/FM interface following the Fert-Levy model \cite{Fert}. Since the thickness of the ferromagnet is usually larger than the charge screening length (1-2 atomic layers), negligible electric field reaches the HM/FM metallic interface. Thus Fert-Levy DMI is not very sensitive to a gate voltage. A second contribution to  DMI arising from the Rashba field at FM/I interface has been recently predicted \cite{Stiles}. Due to its origin, this Rashba DMI contribution is expected to be highly sensitive to applied voltages. 
This tuning of DMI with electric field is paramount to become a strong asset not only for skyrmions but also for other domain wall based devices.

\begin{figure*}
	\centering
	\includegraphics[width=0.9\linewidth]{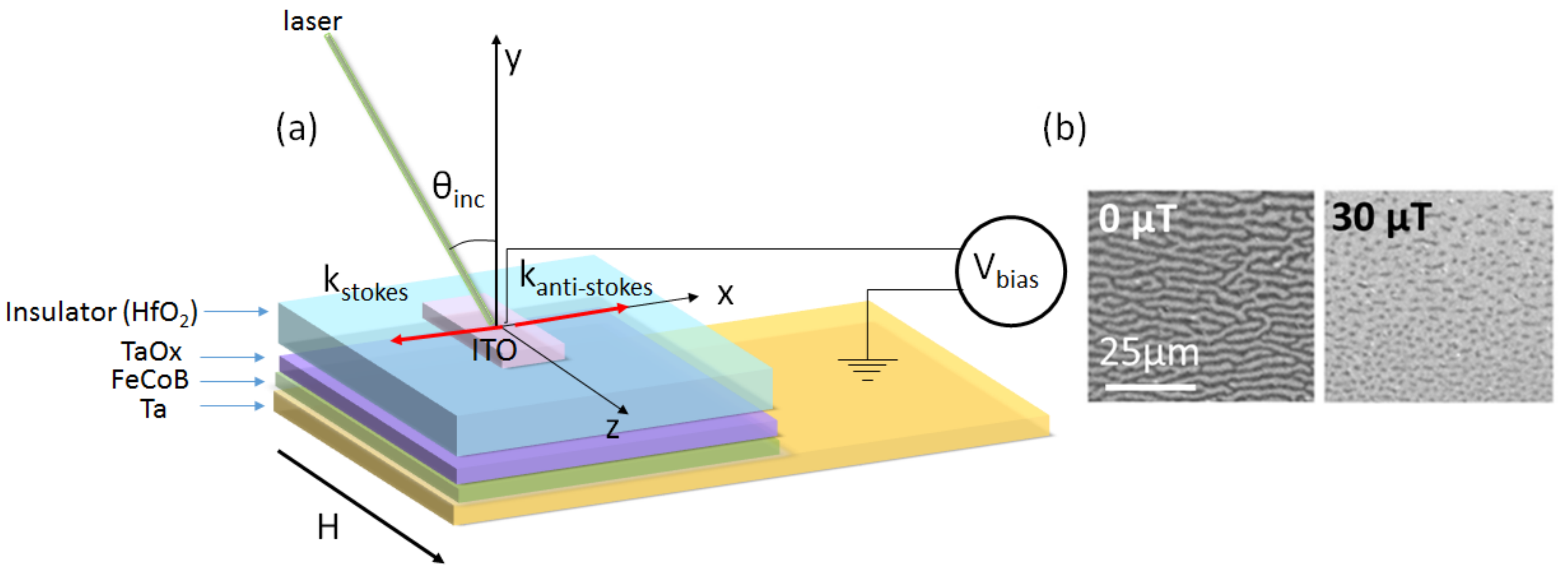}
	\caption{(a) Schematic of the sample and setup for BLS measurement. During the accumulation of the spectra (4h), constant voltage is applied on a $150 \times 800 \mu m^2$ transparent ITO electrode and a magnetic field $H$ is applied to saturate magnetization along $z$. (b) p-MOKE microscopy images of (left) thermally activated demagnetized stripe domains that transform into (right) skyrmionic bubbles on the application of small out of plane magnetic field (30$\mu$T). }
	\label{fig:BLSconfig}
\end{figure*}

\begin{figure*}[h]
\centering
    \includegraphics[width=0.9\linewidth]{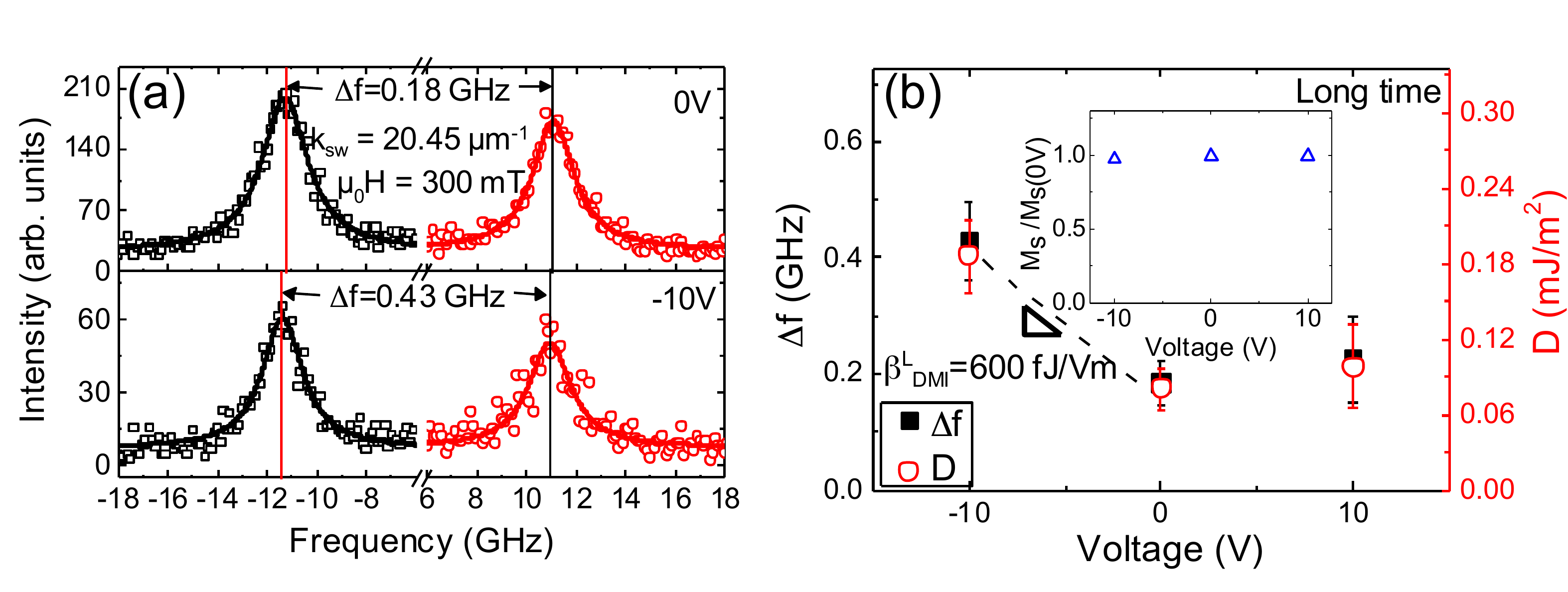}

\caption{(a) BLS spectra (open symbols) measured while applying 0V and -10V and corresponding Lorenztian fits  (solid curve). $f_S$ and $f_{AS}$ correspond to frequencies of the Stokes and the Anti-Stokes peaks and are marked with vertical lines. The shift in frequency $\Delta f=\mid f_S\mid -\mid f_{AS}\mid $ is proportional to the interfacial DMI. It changes by $140\%$ between 0 and -10V. (b) Variation of $\Delta f$ and deduced $D$ as a function of applied voltage. The error bars in $\Delta f$ are calculated from the fits of spectra. To extract $D$, the $M_s$ variation with applied voltage has been taken into account (see inset showing 6\% variation of $M_s$ measured by p-MOKE).}
\label{fig:DMI(V)}
\end{figure*}
In this letter, we report a large variation (130\%) of interfacial DMI by voltage gating using Brillouin Light Spectroscopy (BLS). Moreover, with additional polar-Magneto-Optical-Kerr-Effect (p-MOKE) microscopy, we performed a complete study of long and short time scale effects of electric field on DMI and evaluate their respective efficiencies.  
 
 Our sample consists of Ta(3)/FeCoB(0.9)/TaOx(1) (thicknesses in nm) trilayers deposited by sputtering such that the top Ta is slightly underoxidized (see suppl S1). It is then partially covered by an insulator (HfO$_2$) and a transparent electrode to allow optical observation under voltage (Fig \ref{fig:BLSconfig}a). By p-MOKE microscopy, at zero magnetic field, we observe a spontaneous room temperature thermal demagnetization which leads to stripe domain formation, as shown in Fig \ref{fig:BLSconfig}b. The stripes transform into skyrmionic bubbles on applying a small out of plane magnetic field (30 $\mu$T) (details in suppl S1). The uniform motion of these bubbles under applied current by spin-orbit torques confirms their non trivial topology and hence their skyrmionic nature (see suppl S2): when the domain walls are chiral Néel type, ie. the magnetization always rotates clockwise (or anticlockwise depending on the sign of DMI), the torque  changes sign from one edge of the bubble to the other. As both magnetization inside the wall and torque change sign on opposite edges of the bubble, it results in a motion of the edges in the same direction, which thus leads to a uniform motion of the bubble, without distortion \cite{Jiang}.

To investigate the effect of electric field on DMI, we performed BLS measurements under different applied voltages. 
The schematic setup of the experiment is shown in Fig \ref{fig:BLSconfig}a (see methods for details). BLS spectra measured at 0V and -10 V are represented in Fig \ref{fig:DMI(V)}a. 
The frequency difference $\Delta f$ between Stokes ($f_S$) and anti-Stokes ($f_{AS}$) peaks $\Delta f=\mid f_S\mid -\mid f_{AS}\mid $ is determined from the Lorentzian fits. For a gate voltage of -10V, we observe a significant change of the BLS spectrum, corresponding to 140\% increase of $\Delta f$. By contrast, a very small change in $\Delta f$ is observed for +10V (see Fig \ref{fig:DMI(V)}b). 
Similar non-linearity with voltage has often been reported in studies of electric field effect on interfacial anisotropy \cite{TsymbalEFE,Bauer}. 
Since data acquisition for BLS lasts several hours, the electric field effect measured here corresponds to long time scale phenomena. 
We recover the 0V spectrum (see suppl S3) after the -10V and +10V measurements indicating complete reversibility of the involved mechanisms. 

Interfacial DMI energy $D$ is directly determined from $\Delta f$ using:  $\Delta f=\frac{2\gamma}{\pi M_s}k_{SW}D$  where  $\gamma$ is the  gyromagnetic ratio ($\gamma$/2$\pi$=28.5 GHz/T determined by Ferromagnetic Resonance on 5nm FeCoB films), $k_{SW}$ is the wave vector and $M_s$  the saturation magnetization \cite{Nembach, Di, Ma, BelmePRB2016,BelmePRB2015}. A linear variation of $\Delta f$ as a function of $k_{SW}$  was obtained as expected (see suppl S4); $k_{SW}$ was therefore kept fixed at 20.45 $\mu m^{-1}$ for the measurement of $\Delta f$ as a function of applied voltage. $M_s$ at 0V (1.05 $\times 10^6$ A/m) is measured by Superconducting Quantum Interference Device magnetometry and its variation with voltage is measured from the Kerr signal amplitude of  hysteresis loops observed by p-MOKE (inset of Fig. \ref{fig:DMI(V)}b). Less than 6\% variation of $M_s$ is observed in this voltage range. Taking this variation into account, the evolution of $D$ with voltage is extracted, as represented in Fig. \ref{fig:DMI(V)}b: it varies between  $0.08 \pm 0.01$ and $0.18 \pm 0.03$ $mJ/m^2$, giving a variation of $\Delta D$ = $0.1 \pm 0.04$ $mJ/m^2$. We thus achieve 130\%  variation of DMI in response to an electric field of $E=-170$ $MV/m$.  Taking into account our error bars on DMI estimation, it leads to an increase between 70 and 200\%. It corresponds to a variation of the surface DMI coefficient by  $\Delta D_s=\Delta (D . t)=65 \pm 25$ $fJ/m$, where t is \mbox{FeCoB} thickness (taking into account the magnetically dead layer, t = 0.65 nm, see suppl S1). This variation of $D$ (resp. $D_s$) is three (resp. two) orders of magnitude higher than the only previous experimental observation of voltage induced DMI in thick films \cite{NawaEFDMI} ($\Delta D$ = $40$ $nJ/m^2$ or $\Delta D_s=0.8$ $fJ/m$). 

This variation is also larger than the DMI energy obtained in our sample at 0V. This relatively small DMI value of $0.08$ $mJ/m^2$ is consistent with other studies in similar systems based on Ta/FeCoB, where the top oxide is MgO\cite{Khan, Gross, Ohno}, indicating a small Fert-Levy DMI contribution of Ta/FeCoB interface. It is for instance much smaller than the value around $1-2mJ/m^3$ obtained in Pt/Co/AlOx samples \cite{Pizzini, BelmePRB2015}. By contrast, Rashba-DMI at FeCoB/TaOx interface is likely to exist \cite{Stiles} and to contribute to the total DMI. Our values are similar to the results of Yu et al.\cite{Yu2016, Yu2017}, where the structures are similar with top layer of Ta(0.8-0.9\r{A})/MgO or TaOx.

Moreover, given the short screening length of electric field in a metal, the gate voltage only affects this top FeCoB/TaOx interface. Thus the  strong sensitivity of DMI to gate voltage we have observed suggests that the major contribution to DMI comes from this interface. This result could thus validate the existence of Rashba-DMI  at FM/oxide interfaces and its large sensitivity to electric field.  

To further explore the influence of electric field on  magnetic parameters like anisotropy and exchange, we have studied magnetic domain configuration using p-MOKE microscopy on the same sample at different applied voltages. To be able to compare with the BLS measurements we performed p-MOKE measurements over long time scales (4 hours).
Like in the case of BLS, we recovered the initial magnetic parameters at 0V after the measurements under voltage,  once again confirming the reversibility of the effects. However irreversible changes in the magnetic properties were also observed for higher voltages applied for a long time (see suppl S3).

\begin{figure*}[h]
\centering

    \includegraphics[width=0.9\linewidth]{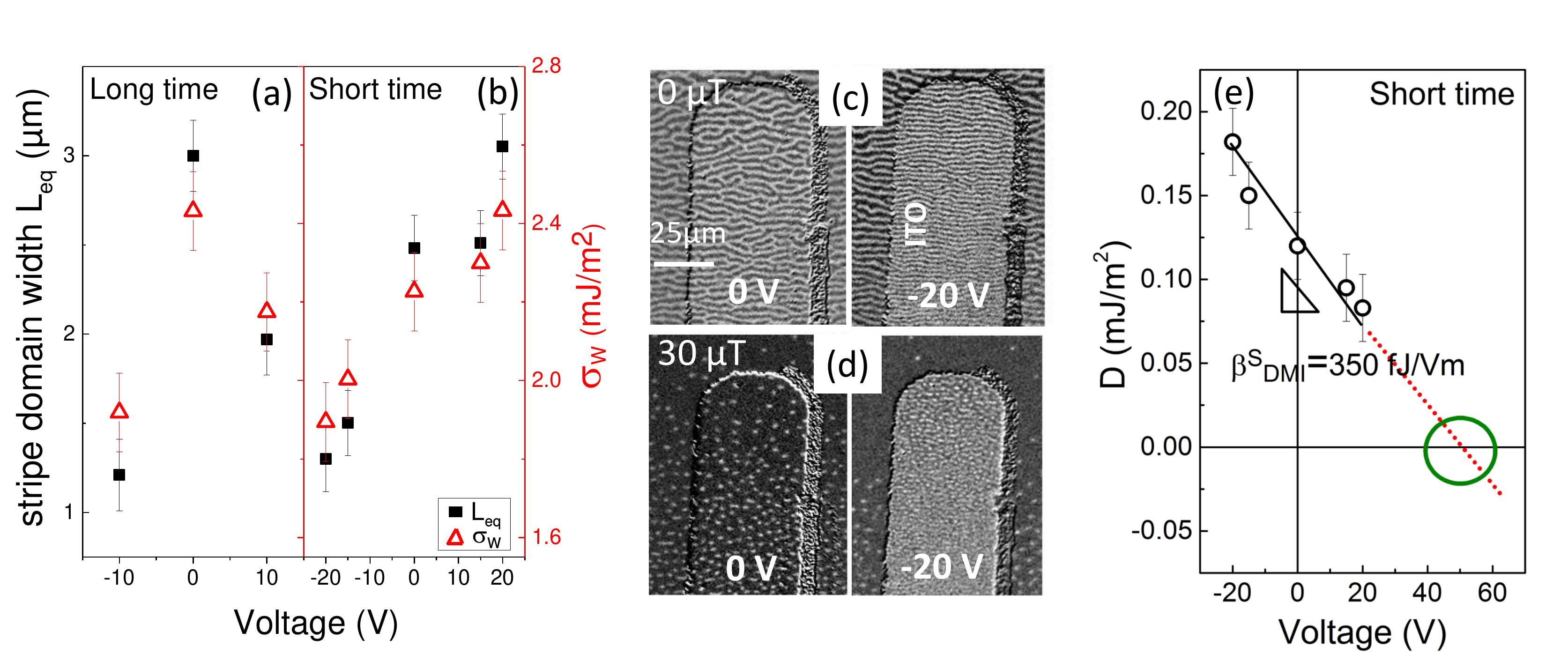}
    
\caption{Measured variation of stripe domain width $L_{eq}$ (squares) and deduced domain wall energy $\sigma_W$ (triangles) as a function of applied voltage for (a) long time scales (4-15h) and (b) short time scales (few mins). As the measurements for the two time scales are performed on two different electrodes, the values at 0V differ slightly. (c,d) p-MOKE images at short time scales in the region of $50 \times 800 \mu m$ ITO electrode at (c) zero and at (d) 30 $\mu T$ out of plane magnetic field. The stripe domain width and the skyrmionic bubbles reduce drastically in size on applying -20V.
(e) Variation of interfacial DMI as a function of  applied voltage inferred for short time scales. A linear extrapolation for larger positive voltages shows that $D$ can be reduced to 0 and  even change sign.}
\label{fig:sigma}
\end{figure*}

We see in Fig. \ref{fig:sigma}c (left) that at zero magnetic field and zero applied voltage, stripe domains are identical both below and outside the electrode. The spontaneous thermal demagnetization results in high nucleation density and domain wall mobility. The stripe domains are hence close to their lowest energy state. By measuring their equilibrium width $L_{eq}$ at different voltages, the corresponding domain wall energy $\sigma_W$ can be extracted using \cite{Marine}: {\center \vspace{-3mm}{$\sigma_W=\frac{\mu_0M_s^2t}{\pi}ln(\frac{L_{eq}}{\alpha t})$ }} 

with $t$ the ferromagnetic film thickness, and $\alpha$ a constant taken equal to 0.955. This model is valid if the domain size is much larger than the domain wall width and the ferromagnetic film thickness smaller than the characteristic dipolar length \cite{Kaplan}, which is the case in our system. At negative voltages, $L_{eq}$ undergoes a strong reduction in size and the stripes become closely spaced as seen in Fig. \ref{fig:sigma}c (right) (from 2.4 $\mu m$ at 0V to 1.1 $\mu m$ at -20V). Taking into account the small variation of $M_s$ in this voltage range, we deduce $\sigma_{W}$: it decreases strongly for negative voltage whereas a weak decrease is observed for positive voltage as shown in Fig \ref{fig:sigma}a.

\begin{figure*}[h]
\centering

    \includegraphics[width=0.9\linewidth]{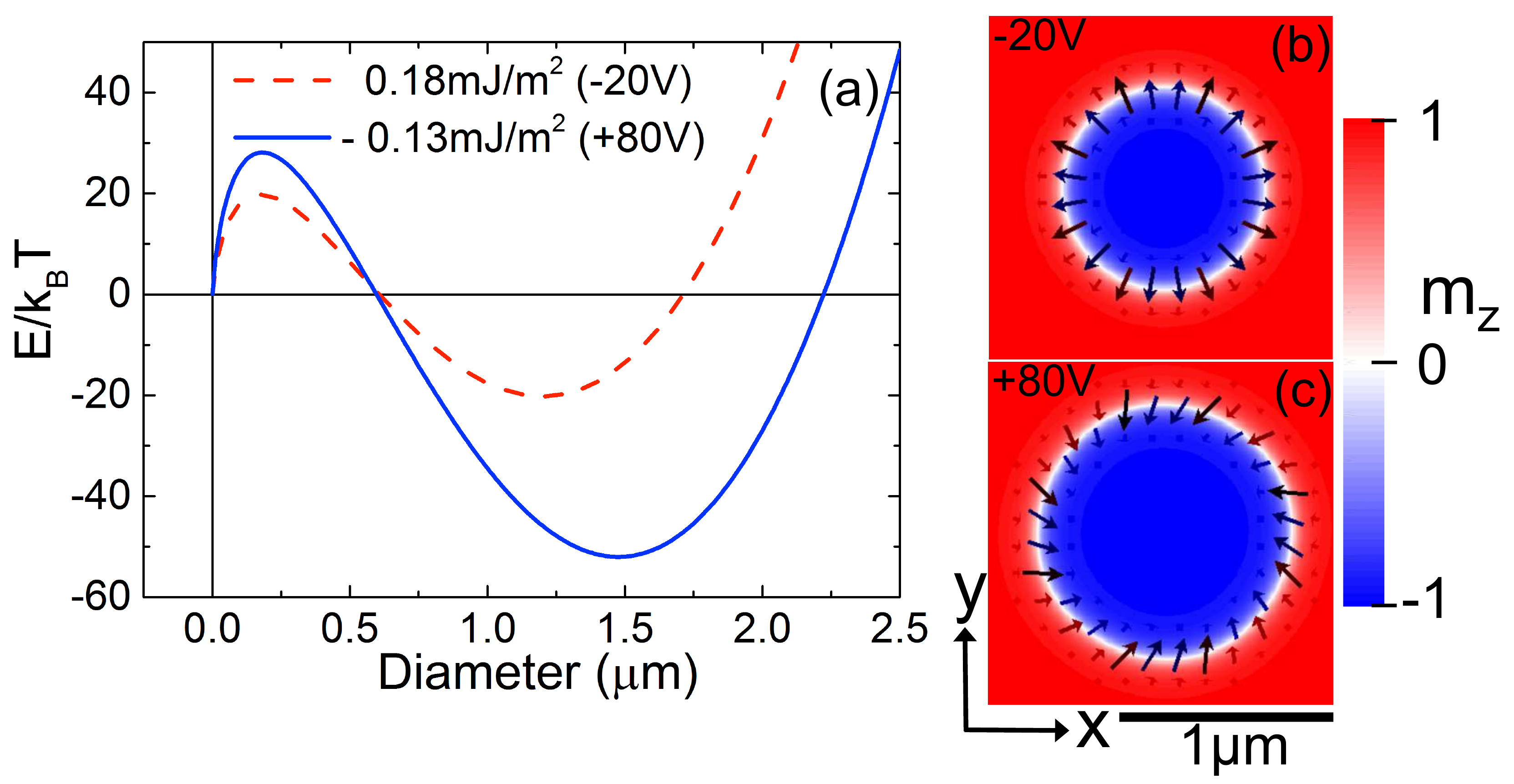}
    
\caption{(a) Analytical model: variation of the energy difference between the bubble state and the saturated state as a function of the bubble diameter. Dipolar, anisotropy,  exchange, Zeeman and DMI energies have been taken into account. The magnetic parameters used for the two curves are close to the ones measured  for the -20V case (short time) and extrapolated from the voltage variation of these parameters for the +80V case. In the -20V (resp. +80V) case, DMI is positive (resp. negative) and the energy is calculated for a circular bubble with the corresponding lower energy chirality. In both cases, a stable bubble is obtained.
(b,c) Micromagnetic simulations : distribution of magnetization of skyrmionic bubbles obtained in a 6 $\mu$m dot with  parameters corresponding to the ones of the analytical model for -20V (b) and +80V (c). In the case of -20V, 
the spins point outwards while in the +80V case, with opposite DMI sign, 
an opposite chirality  is stabilized. In this latter case however, the domain walls are not completely N\'eel type . The parameters for these calculations are given in supplementary S7.
}
\label{fig:simul}
\end{figure*}

The domain wall energy in the presence of an interfacial DMI is given by $\sigma_W=4\sqrt{AK}-\pi \mid D\mid$ \cite{Woo,Heide} with $K$ the effective anisotropy ($K=K_s/t -0.5 \mu_0 M_s^2$ where $K_s$ is the surface anisotropy) and $A$ the exchange stiffness. By combining BLS and p-MOKE results on  DMI and $\sigma_W$  respectively, we deduce that $\sqrt{AK}$ at these long time scales varies by less than 10\% in this voltage range.   

In  view of  applications, it is crucial to study the dynamic change of interfacial DMI with voltage at shorter time scales. We hence performed p-MOKE microscopy by applying voltages for a few minutes. At these durations, contrary to long time scale observations, $L_{eq}$ and $\sigma_W$ vary monotonically with voltage, as represented in Fig \ref{fig:sigma} b. This highlights the difference in the electric field effect mechanisms for the two time scales. This monotonic trend was qualitatively observed for short time scales on all electrodes that we have measured on this sample and also on an equivalent one from another wafer. 

In order to extract DMI variation under voltage at short time scales we also measured the variations of $H_k$ and $M_s$ (see suppl S5) which were found to be below 12\% and 6\% respectively in the range of $\pm$ 20V. They both decrease (increase) with negative (positive) voltages which is in agreement with other studies\cite{MatsuEFE}. Since we show that the variation of $M_s$ is small, we consider exchange stiffness $A$ to be constant in this voltage range (see supp S6).  
The DMI variation with voltage at short time scale is then extracted, as plotted in Fig \ref{fig:sigma}e.

The electric field efficiency at negatives voltages $\beta_{DMI}=\Delta D/ E$  is twice larger for  long ($\beta^L_{DMI}=600 fJ/Vm$) as compared to short ($\beta^S_{DMI}=350 fJ/Vm$) time scales. Both are however much larger than  theoretical prediction\cite{Mairarxiv} ($\beta_{DMI}=26 fJ/Vm$) and the previous experimental demonstration in thick film\cite{NawaEFDMI} ($\beta_{DMI}=3 fJ/Vm$). The qualitative difference between short and long time scale results could be ascribed to the different physical mechanisms involved. We have also evaluated $\beta_{K_s}(=\Delta K_s/E)$ and we obtain values around $100$ (short-time scale) and $200 fJ/Vm$ (long time scale), which are a bit smaller but of the same order of magnitude as $\beta_{DMI}$. This is expected as both Rashba-DMI and surface anisotropy find their origin in spin-orbit coupling. In fact the 12\% variation of anisotropy field corresponds to a similar variation of $K_s$ in absolute values as compared to the variation of DMI; but these variations presented in relative values are very different since the absolute value of $K_s$ is high whereas the absolute value of DMI is low.

 We now propose tentative explanations for these behaviors.
 For short time measurements, negative electric field would add up to the Rashba field (thus increasing the DMI), while a positive one would partially compensate it \cite{Barnes}. Therefore we propose that the large DMI sensitivity to gate voltage can be explained by the Rashba-DMI contribution originating from the FeCoB/TaOx interface.
 
We further suggest that the stronger amplification of DMI at negative voltage for long time scale is related to the additional effect of oxygen migration towards the \mbox{FeCoB/TaOx} interface. Since the FM/I interface is slightly underoxidized (see suppl S1), this ion migration results in a more optimally oxidized interface reinforcing the Rashba-field.
However, for positive voltages, the DMI evolution at long time scale is more complex to analyze. The shift of the oxidation front away from the FM film, may strongly alter both the Rashba field and any charge effect via changes in crystalline and orbital structures. Similarly, the voltage induced variation of anisotropy due to ion migration was observed to be asymmetric in other systems and strongly dependent on the location of the oxygen front (described in the supplementary material of Bauer et al.\cite{Bauer}). As both the anisotropy and  DMI stem from spin orbit coupling at this interface, this scenario could explain a similar behavior versus voltage for DMI in our system.

To study the effect of voltage tuning of DMI on skyrmionic bubbles size and density for short time scales, we use p-MOKE (Fig \ref{fig:sigma}d): for negative voltages (-20V) the skyrmionic bubbles under the electrode show a higher density and a 50\% smaller size as compared to 0V. These observations follow the evolution of the stripe domain width at short time scale (Fig \ref{fig:sigma}c). Such a behavior was previously shown by us in Pt/Co/AlOx system \cite{Marine}. 
However, in the latter case, a strong variation of anisotropy and saturation magnetization with electric field was fundamentally responsible for the change in size of skyrmionic bubbles. By contrast, in the present case of Ta/FeCoB/TaOx, the primary cause of this effect is a huge voltage induced DMI variation whereas $M_s$ and $K$ vary in smaller proportion.

In order to further explore  the DMI tuning in wider voltage range, we extrapolate to higher positive voltages the monotonic variation of $D$ obtained for short time scale: we expect a complete cancellation or even a sign reversal of $D$ at around +40V. It corresponds to 670 MV/m and is below the typical breakdown voltage of oxides in spintronic devices (typically 600-900 MV/m for HfO$_2$ \cite{Conley} or 1000-1500 MV/m for MgO \cite{Amara}). Theoretical studies shows that  DMI sign reversal is possible by playing on the oxygen coverage of the interface \cite{Belabbes}. Additionally, an approach through Ar$^+$ irradiation has also been employed to locally control the sign of DMI \cite{Balk}. Here we put forward the possibility of controlling the DMI sign both locally and dynamically through voltage gating. 

To further investigate the effect of this DMI sign reversal, we performed both analytical calculations and micromagnetic simulations. First, we have used the analytical model of isolated bubble developed in  ref \cite{Marine} taking into account the dipolar, exchange, anisotropy, Zeeman and DMI energies (see Fig \ref{fig:simul}a). The input magnetic parameters and their variation with voltage are similar to the ones we have  measured or extracted  from the short time scale experiments and are extrapolated to +80V (see suppl. S7). In the -20V case, D is equal to  0.18 $mJ/m^2$. For the +80V case, the extrapolation of DMI (fig \ref{fig:sigma}e) leads to a negative value of similar magnitude as the -20V experimental case (D= -0.13 $mJ/m^2$). We show that, for the two sets of parameters corresponding to -20V and +80V, ie. to D=+0.18 and -0.13 $mJ/m^2$ respectively, stable skyrmionic bubbles of micrometer size and opposite chirality can be obtained. 

In the analytical model, the chirality is introduced by calculating the average domain wall energy as previously explained.  Any inhomogeneities in domain wall structure or any non circular shape domain is thus not taken into account. Consequently, in order to access the internal domain wall structure and the exact domain shape, we further performed micromagnetic simulations using Mumax3 \cite{Vansteenkiste} (see suppl S7) with similar magnetic parameters (Fig \ref{fig:simul}b,c). For the -20V state, a circular skyrmionic bubble of 1.1 $\mu$m diameter with 
outward spins within the domain wall is stabilized (N\'eel domain wall). The obtained size is very close to experimental values. For the +80V state, a circular skyrmionic bubble state is also obtained but with opposite chirality, ie. the spins point inward. The diameter in this latter case is slightly increased, consistently with the analytical calculations. The bubble diameter is 1.4 $\mu$m and the domain wall is not exactly N\'eel type: in fact the magnetization at the center of the domain wall  presents a constant angle of 26\ensuremath{^\circ} with respect to the radial direction. Hence, it is rather a Dzyaloshinskii domain wall \cite{Thiaville}. This is probably due to the fact that the DMI amplitude is close to the critical DMI necessary to obtain pure N\'eel walls. We could not further increase  the positive voltage as it led, with our material parameters, to negative domain wall energies and thus to spin spiral state. However, these simulations, combined with the analytical model results, show that a chirality switch controlled by gate voltage can be designed in this type of systems. Further material engineering to adjust DMI, $M_s$ and $K_s$ variations under voltage and also Rashba-DMI contribution would be necessary to optimize this behavior for applications.

To conclude, we have measured 130\% variation of DMI coefficient under voltage using Brillouin Light Spectroscopy in a Ta/FeCoB/TaOx trilayer. A monotonic variation of DMI with voltage is further observed at shorter  time scales by p-MOKE microscopy. The huge electric field effect on DMI might be explained by considering the fact that the main contribution to DMI stems from the interface with the oxide layer (Rashba-DMI), and is therefore directly sensitive to the gate voltage. We further anticipate a sign reversal of DMI at higher voltages enabling a chirality switch. This result may ultimately lead to a dynamic control of the skyrmion properties, for electrically programmable skyrmion based memory and logic devices.

\vspace{0.5cm}
\section*{{\large Methods}}

\paragraph*{Samples:}
The Ta(3)/FeCoB(0.9)/TaOx(1) (thicknesses in nm) trilayers have been deposited by magnetron sputtering on thermally oxidized Si wafer. To observe electric field effect on these skyrmionic bubbles, a thick (60nm) insulator  of HfO$_2$ was deposited by Atomic Layer Deposition to prevent current flow through the barrier over wide lateral sizes. This is followed by sputtering of transparent Indium Tin Oxyde (ITO) film which is further patterned into electrodes of several tens $\mu m$ lateral size.  This electrode  allows performing both BLS and p-MOKE microscopy under an applied voltage. 

\paragraph*{BLS:}
The magnetic field is applied perpendicular to the incidence plane allowing the probing of spin waves propagating in the plane perpendicular to the applied magnetic field in the Damon-Eshbach (DE) geometry, as represented in fig \ref{fig:BLSconfig}a. The applied magnetic field is above the saturation field of the sample as determined from magnetometry loops. The frequency shift in the Stokes and Anti-Stokes resonances is analyzed using a 2x3 Fabry-Perot interferometer (3-300 GHz) and is calculated by $\Delta f=\mid f_S\mid -\mid f_{AS}\mid $. Calibrations were done to remove any offset \cite{Bouloussa}. The laser spot of $100\mu m$ in diameter was focused on the electrode which is connected to the voltage source. The  spin wave vector is given by $k_{sw} =4\pi sin(\theta_{inc})/\lambda$ where $\theta_{inc}=60^{\circ}$ is the angle of incidence and $\lambda=532$ nm the wavelength of the incident laser. We have measured spectra for 0, -10V and +10V applied voltage for a counting time of 4 hours.
 
\section*{Supporting information}
Optimization of material parameters to get skyrmionic bubbles. Motion under current (video).
Time scale effects and reversibility of measurements. Evolution of $\Delta f$ with $k_{SW}$. Evolution of $M_s$ and $H_k$ with gate voltage. Measured and extracted parameters on different time scales. Analytical model of isolated bubble and micromagnetic simulations.

\begin{acknowledgement}

The authors acknowledge funding by the French ANR (contract ELECSPIN ANR-16-CE24-0018), by the University Grenoble Alpes (project Automag2D), by Nanosciences fondation and fruitful discussions with Mihai Miron.

\end{acknowledgement}

The authors declare no competing financial interest.

\end{document}